\RequirePackage{tikz}

\documentclass[pdflatex,sn-mathphys]{sn-jnl}

\usepackage[utf8]{inputenc}
\usepackage[T1]{fontenc}
\usepackage{xcolor}
\usepackage{comment}
\usepackage{colortbl}
\usepackage{url}
\usepackage{tabularx}
\usepackage{booktabs}
\usepackage{pifont}
\usepackage{xspace}
\usepackage{tikz}
\usepackage{pgfplots}
\usepackage{pgfplotstable}
\pgfplotsset{compat=1.14}
\usetikzlibrary{fit,positioning,calc,shapes,backgrounds,external}
\usepackage{hyphenat}
\usepackage{balance}
\usepackage{multirow}
\usepackage{nicefrac}
\usepackage{enumitem}
\usepackage{algorithm}
\usepackage{algpseudocode}
\usepackage{subcaption}
\usetikzlibrary{matrix}
\usetikzlibrary{patterns}
\usepgfplotslibrary{groupplots}
\usepackage{wrapfig}
\usepackage{amsthm}
\usepackage{amsfonts}
\usepackage{amsmath}
\usepackage{amssymb}
\usepackage[normalem]{ulem}
\usepackage{graphicx}
\usepackage{textcomp}
\usepackage{rotating}
\usepackage{mdframed}
\usepackage[noindentafter]{titlesec}
\usepackage{bm}

\newcommand{\shortauthor}{Moradan et al.}
\jyear{2023}%

\theoremstyle{thmstyleone}%
\theoremstyle{thmstyletwo}%
\theoremstyle{thmstylethree}%

\DeclareMathOperator{\tr}{Tr}
\DeclareMathOperator*{\argmax}{\arg\:\max}

\DeclareMathOperator{\Diag}{dg}

\newcommand{\spara}[1]{\smallskip\noindent{\bf #1}}
\newcommand{\mpara}[1]{\medskip\noindent{\bf #1}}
\newcommand{\para}[1]{\noindent{\bf #1}}

\algnewcommand\Params{\item[\textbf{Params:}]}

\definecolor{cycle1}{RGB}{228, 26, 28}
\definecolor{cycle2}{RGB}{55, 126, 184}
\definecolor{cycle3}{RGB}{77, 175, 74}
\definecolor{cycle4}{RGB}{152, 78, 163}
\definecolor{cycle5}{RGB}{255, 127, 0}
\definecolor{cycle6}{RGB}{153, 153, 153}
\definecolor{cycle7}{RGB}{166, 86, 40}
\definecolor{cycle8}{RGB}{247, 129, 191}

\definecolor{eyecancerpink}{RGB}{252, 15, 192}

\newtheorem*{problem*}{Problem}
\surroundwithmdframed [%
  innertopmargin = -13pt,
  innerleftmargin = 5pt,
  innerrightmargin = 5pt,
  backgroundcolor = gray!12,%
  startinnercode={\baselineskip=1.44\baselineskip}
] {problem*}

\newcommand*{\NPhard}{$\mathbf{NP}$-hard\xspace}

\newcommand{\cmark}{\textcolor{black}{\ding{52}}} 

\newcommand{\win}{\textcolor{cycle3}{$\uparrow$}} 
\newcommand{\lose}{\textcolor{cycle1}{$\downarrow$}}

\newcommand{\xmark}{\textcolor{gray}{\ding{56}}}

\newcommand{\cvec}{\mathbf{c}}
\newcommand{\xvec}{\mathbf{x}}

\newcommand{\dvec}{\mathbf{d}}

\newcommand{\yvec}{\mathbf{y}}
\newcommand{\amat}{\mathbf{A}}
\newcommand{\bmat}{\mathbf{B}}

\newcommand{\cmat}{\mathbf{C}}
\newcommand{\wmat}{\mathbf{W}}
\newcommand{\xmat}{\mathbf{X}}
\newcommand{\dmat}{\mathbf{D}}

\newcommand{\adj}{\amat}

\newcommand{\eye}{\mathbf{I}}

\newcommand{\reals}{\mathbb{R}}

\newcommand{\dm}[1]{#1}
\newcommand{\ia}[1]{#1}

\newcommand{\amw}[1]{{\color{black}#1}}

\newcommand{\dmw}[1]{#1}

\newcommand{\dmk}[1]{{#1}}

\newcommand{\dmi}[1]{{#1}}

\newcommand{\dml}[1]{{\color{black}#1}}
\newcommand{\revc}{{\cellcolor{white!20}}}
\newcommand{\rev}[1]{#1}

\newcommand{\dme}[1]{#1}

\newcommand{\revm}[1]{#1} 

\newcommand{\comm}[1]{}

\newcommand*{\Gr}{\mathcal{G}}
\newcommand{\V}{\mathcal{V}}
\newcommand{\E}{\mathcal{E}}

\newcommand{\dg}{d}
\newcommand{\dgvec}{\dvec}
\newcommand{\dgmat}{\dmat}

\newcommand{\nn}{n}
\newcommand{\na}{l}

\newcommand{\Cl}{\mathcal{C}}

\newcommand{\cora}{Cora\xspace}
\newcommand{\citse}{Citeseer\xspace}
\newcommand{\pubmed}{Pubmed\xspace}
\newcommand{\amzph}{Amz-Pho\xspace}
\newcommand{\amzpc}{Amz-PC\xspace}
\newcommand{\coacs}{CoA-CS\xspace}
\newcommand{\coaphy}{CoA-Phy\xspace}
\newcommand{\fb}{Fb\xspace}
\newcommand{\eng}{Eng}

\newcommand{\ourmethod}{UCoDe\xspace}
\newcommand{\thiswork}{\ourmethod}
\newcommand{\kmeans}{$k$-means\xspace}
\newcommand{\louvain}{Louvain\xspace}
\newcommand{\dgi}{DGI$^k$\xspace}
\newcommand{\mincut}{MinCut\xspace}
\newcommand{\nocd}{NOCD\xspace}
\newcommand{\dmon}{DMoN\xspace}
\newcommand{\ourmethodk}{UCoDe$^k$}
\newcommand{\dcrn}{DCRN\xspace}
\begin{document}

\title[\ourmethod]{\ourmethod: Unified Community Detection with Graph Convolutional Networks}


\author*[1]{\fnm{Atefeh} \sur{Moradan}}\email{atefeh.moradan@cs.au.dk}

\author[1]{\fnm{Andrew} \sur{Draganov}}\email{draganovandrew@cs.au.dk}

\author[1]{\fnm{Davide} \sur{Mottin}}\email{davide@cs.au.dk}

\author[1]{\fnm{Ira} \sur{Assent}}\email{ira@cs.au.dk}

\affil*[1]{\orgdiv{Department of Computer Science}, \orgname{Aarhus University}, \orgaddress{\street{Aabogade 34}, \city{Aarhus}, \postcode{8200}, \country{Denmark}}}




\abstract{
\begin{abstract}





\dml{Community detection finds \dmk{homogeneous groups of nodes in a graph}. Existing approaches  either partition the graph into disjoint, \emph{non-overlapping}, communities, or determine only \emph{overlapping} communities. \dmk{To date}, no method supports both detections of overlapping and non-overlapping communities.} 
We propose \ourmethod, a \emph{unified} method for community detection in attributed graphs \dml{that detects both overlapping and non-overlapping communities by means of a novel contrastive loss that captures node similarity on a macro-scale. }
\dmk{Our thorough experimental assessment on real data shows that, regardless of the data distribution, our method is either the top performer or among the top performers in both overlapping and non-overlapping detection without burdensome hyper-parameter tuning. 
}

\end{abstract}}




\maketitle

\section{Introduction}\label{sec:introduction}

\ia{Community detection~\cite{fortunato2010community} is the problem of identifying sets of nodes in a graph that share common characteristics. 
In social networks, community detection identifies groups of individuals who participate in joint activities (e.g. sports clubs) or having similar preferences~\cite{perozzi2014deepwalk}; in biological networks, communities represent proteins that contribute to a specific disease~\cite{mall2017adaptive}.
\dml{Such networks include information in node attributes that may be helpful when identifying similarities (e.g., the age of a person).}
\dmw{However, these attributes are typically not considered} by traditional community detection methods, such as spectral clustering~\cite{shi2000normalized}, modularity maximization~\cite{newman2006modularity}, or more recent graph embeddings~\cite{cai2018comprehensive}, making them ill-suited for detecting node communities in \dmw{attributed} graphs.}

\ia{In recent years, graph neural networks (GNNs)~\cite{kipf2017GCN, velivckovic2018deep, hamilton2017inductive, bronstein2017geometric, battaglia2018relational} have shown superior performance in a number of supervised tasks on graphs, especially link prediction, node classification, and graph classification. 
GNNs popularity stems from their aptitude to capture complex relationships in networks, typically by means of propagating node attributes and features to neighboring nodes by a message-passing process~\cite{battaglia2018relational}. \dml{These are typically accompanied by 
\emph{graph pooling}~\cite{bruna2014spectral,bianchi2020spectral,lee2019self}, which aggregates multiple nodes into higher-level representations to reduce the number of parameters of the neural network.}}

\ia{GNNs have propelled advancements in supervised tasks\dmw{; yet on} unsupervised tasks such as community detection, \dmw{GNNs have not yet received the same attention.}
Most existing GNN methods do not directly optimize for community detection but achieve the objective indirectly.
Unsupervised GNNs, such as the popular Deep Graph Infomax (DGI)~\cite{velivckovic2018deep}, find node representations that, in a second step, need to be subjected to a clustering algorithm, such as the widely used k-means, to actually obtain communities. 
} 

\ia{Recently, a few methods propose GNNs that \emph{explicitly} optimize for community detection. 
\dmw{GNNs for \emph{non-overlapping} community detection}
either optimize \dmw{for} a \emph{single score} or \emph{combine several scores}. 
\dmw{Single score} methods revisit traditional measures such as min-cut~\cite{bianchi2020spectral} and modularity~\cite{tsitsulin2020graph} objectives to return node-community probabilities. 
Combined score methods~\cite{zhang2019attributed, zhang2020commdgi} integrate multiple different objectives. \dmw{These methods outperform single score methods in non-overlapping community detection, but} require substantial tuning to the dataset at hand and are typically less robust and less interpretable than their single objective counterparts.  

\dmw{Non-overlapping community detection aims at returning a single community assignment for each node. As such, they are ill-suited for overlapping community detection.} 
\dmw{NOCD~\cite{shchur2019overlapping} is, at the time of writing this paper, the only GNN that optimizes for \emph{overlapping community} detection. In particular, NOCD finds communities that maximize the probability of recovering the graph structure. Yet, this approach constrains the community structure to be overlapping and thus does not capture non-overlapping communities.
In conclusion, to date, no GNN detects both overlapping and non-overlapping communities. 
}
}




\ia{\para{Contributions.} \textbf{(1)} We introduce a new GNN method, \ourmethod, for community detection on graphs. We devise a simple effective single score model which leverages state-of-the-art representations;  \textbf{(2)} \ourmethod features a novel contrastive loss function that promotes both overlapping and non-overlapping communities, thus being the first approach to achieve competitive results across these tasks with a single model. \textbf{(3)} We perform extensive experiments on real data, showing that our method outperforms single-objective methods without the need for extensive parameter tuning, achieving quality on par with more complex combined scores. 
}

\section{Related Work}\label{sec:related-work}
Before delving into our solution, we provide an overview of the literature on \dmw{community detection, graph neural networks, graph pooling, and graph embeddings}. Table~\ref{tbl:relatedwork} provides a summary of the characteristics of the most important work in the area, highlighting core properties of the methods, such as their ability to capture overlapping and non-overlapping communities and whether they achieve their results in an unsupervised manner with a single score approach. 

\subsection{Traditional community detection}

\ia{Community detection has a long history in graph analysis~\cite{fortunato2010community} with applications across the natural sciences. 
There are two main categories of community detection: \textit{non-overlapping} community detection, also called partitioning, which seeks an assignment of each node to exactly one community; \textit{overlapping} community detection, seeking a soft-assignment of nodes into potentially multiple communities. 
A community detection algorithm optimizes a score that describes the cohesiveness of nodes in the community with respect to the rest of the nodes. 
A number of scores and methods have been proposed based on the graph structure, such as spectral clustering for min-cut~\cite{shi2000normalized}, Louvain's method for modularity~\cite{newman2006modularity}, and the Girvan-Newman algorithm for betweenness~\cite{Girvan_2002/betweeness}. \rev{Other works extend such methods by incorporating node features into the graph analysis~\cite{yang2013community}.}

Overlapping community detection is often approached using algorithms similar to the Expectation-Maximization algorithm for soft-clustering~\cite{dempster1977maximum} where each point is a distribution over the clusters. Similarly, AGM~\cite{yang2012community} and BigCLAM~\cite{yang2013overlapping} formulate the community detection problem as finding soft assignments to communities that maximize the model likelihood. \rev{Other traditional methods find overlapping communities by removing high-betweenness edges~\cite{gregory2007algorithm} or by propagating label information~\cite{gregory2010finding}. Lastly,} {EPM}~\cite{zhou2015infinite} fits a Bernoulli–Poisson model, {SNMF}~\cite{SNMF2010} and {CDE}~\cite{CDE2018} use non-negative matrix \dmw{factorization.}
}

\subsection{Graph Neural Networks for community detection}

\begin{table}[htb]
\scriptsize
\setlength{\tabcolsep}{2pt}
\newcolumntype{C}{>{\centering\footnotesize\arraybackslash}X}
\begin{tabularx} {\linewidth}{p{1.7cm}CCCCCCC}
& \multicolumn{4}{c}{\textbf{loss function}} & \multicolumn{3}{c}{\textbf{model}}\\
\cmidrule(lr){2-5}\cmidrule(lr){6-8}
\bf{method} & Unsuperv. &  Modularity & Contrastive & Single score & Non-overlap & Overlap & Trainable   \\ 
\midrule
DGI~\cite{velivckovic2018deep}      & 	\cmark & \xmark & \cmark & \cmark & \cmark & \xmark & \cmark \\
Graclus~\cite{dhillon2007weighted}	& 	\xmark & \xmark & \xmark & \cmark & \cmark & \xmark & \xmark \\
DiffPool~\cite{Ying2018diffpool}    & 	\cmark & \xmark & \xmark & \cmark & \cmark & \xmark & \xmark \\
MinCut~\cite{bianchi2020spectral}   & 	\cmark & \xmark & \xmark & \cmark & \cmark & \xmark & \cmark \\
CommDGI~\cite{zhang2020commdgi}	    &   \cmark & \cmark & \xmark & \xmark & \cmark & \xmark & \cmark \\
\dmon~\cite{tsitsulin2020graph}	    &   \cmark & \cmark & \xmark & \cmark & \cmark & \xmark & \cmark \\
AGC~\cite{zhang2019attributed}	    & 	\cmark & \xmark & \xmark & \cmark & \cmark & \xmark & \cmark \\
\nocd~\cite{shchur2019overlapping}	& 	\cmark & \xmark & \xmark & \cmark & \xmark & \cmark & \cmark \\
\dcrn~\cite{liu2022deep}	& 	\cmark & \xmark & \xmark & \xmark & \cmark & \xmark & \cmark \\

\midrule
\textbf{\ourmethod} & \cmark & \cmark{} & \cmark{} &  \cmark & \cmark{}  & \cmark  & \cmark{}\\
\bottomrule
\end{tabularx}

\caption{Related work in terms of present (\cmark) and absent (\xmark) properties. }
\label{tbl:relatedwork}
\vspace{-4mm}
\end{table}

GNNs~\cite{wu2020comprehensive} are a family of parametric models that learn node representations by aggregating features over the graph's structure. \dme{GNNs exhibit state-of-the-art performance in supervised tasks, such as link prediction, node and graph classification.}

Popular GNN models include spectral GNNs~\cite{bronstein2017geometric, defferrard2016convolutional}, GCNs~\cite{hamilton2017inductive,kipf2017GCN}, graph autoencoders (GAEs)~\cite{kipf2016variational}, graph isomorphism networks~\cite{xu2018powerful}, and Deep Graph Infomax (DGI)~\cite{velivckovic2018deep}. These models compute node features in an unsupervised manner if equipped with a reconstruction loss. A clustering algorithm, such as \kmeans, can cluster the node features to return communities. Since there is no coupling between such GNN model objectives and the clustering algorithm, the resulting communities may not accurately represent all groups in the graph.  

\spara{GNNs for community detection.} \noindent Some GNNs directly optimize for non-overlapping community detection with community-wise loss functions. \emph{Single objective} approaches propose variations of traditional cohesiveness scores, such as min-cut~\cite{bianchi2020spectral} and modularity~\cite{tsitsulin2020graph}. \dmw{Yet, single-objective methods inherit the limitations of the score they aim to optimize, providing community memberships that are subject to the loss objective's definition of community.} 

CommDGI~\cite{zhang2020commdgi} proposes a \emph{combined objective} as a linear combination of three objectives, the DGI objective~\cite{velivckovic2018deep}, modularity, and mutual information. CommDGI's combined objective overcomes the limitations of the single score methods but requires extensive parameter tuning for proper results. \rev{Similarly, recent multi-objective methods operate on the pairwise correlation matrix~\cite{liu2022deep}, unsupervised contrastive relations~\cite{park2022cgc}, KL-divergence between clusters~\cite{zhao2021graph}~\cite{bo2020structural}, and structured encodings~\cite{he2021community}. \dme{These methods, besides employing complex combined objectives, often require initialization with elaborate pre-trained models~\cite{liu2022deep,zhao2021graph,bo2020structural}, running \kmeans either in the computation of the embeddings~\cite{liu2022deep}, in each epoch~\cite{sun2021graph}, or as an initialization step~\cite{bo2020structural}, and hyperparameter tuning \emph{for each dataset}~\cite{liu2022deep,park2022cgc,zhao2021graph,bo2020structural}. In contrast, our model uses \emph{the same hyperparameters} for all datasets, devises a single-objective contrastive loss, requires no sophisticated initialization, and detects communities without the need to run \kmeans. Nevertheless, in our evaluation, we also compare with \dcrn~\cite{liu2022deep}, the most recent of such combined objective methods. }




\dmw{While models like \dmon~\cite{tsitsulin2020graph} return soft community assignments through a softmax output layer, both single and combined objective methods explicitly penalize overlap among communities. 

\nocd~\cite{shchur2019overlapping} proposes an overlapping community detection loss that maximizes the likelihood of Bernoulli-Poisson models~\cite{shchur2019overlapping}. \nocd achieves competitive results on overlapping community detection but cannot directly detect non-overlapping communities.}
}

\subsection{Graph Pooling}
\ia{Graph pooling~\cite{bruna2014spectral,bianchi2020spectral,lee2019self} is an operation that aggregates nodes so as to learn summarized representations. 
The purpose of graph pooling is to remove redundant information and reduce the number of parameters of the GNN. 

\emph{Model-free pooling} coarsens the graph structure by aggregating nodes without considering the node attributes. Graclus~\cite{dhillon2007weighted} revisits max-pooling to aggregate similar nodes in a hierarchical fashion. SAGPool~\cite{lee2019self} proposes a self-attention layer to reweigh nodes in the graph. Model-free approaches act as layers in the network and do not provide communities as output. 

\emph{Model-based pooling} learns coarsening operators through a differentiable loss function. DiffPool~\cite{Ying2018diffpool} learns a hierarchical clustering assignment of the graph for supervised graph classification. Top-K pooling~\cite{Ga02019TopK} trains an autoencoder that assigns a score to each node; the pooling phase retains the k nodes with the highest score. Yet, these methods do not explicitly optimize for cluster assignments resulting in substandard communities~\cite{bianchi2020spectral}. 

MinCutPool~\cite{bianchi2020spectral}, although a pooling technique, returns community assignments by optimizing the min-cut objective of spectral clustering~\cite{shi2000normalized}. MinCutPool does not require eigendecomposition of the Laplacian matrix and instead propagates node attributes over the GNN. 
}

\subsection{Node embedding methods}
\ia{Node embeddings~\cite{cai2018comprehensive,chami2020machine} learn node representations of the graph structure in an unsupervised manner with shallow neural networks~\cite{perozzi2014deepwalk, tang2015line}, autoencoders~\cite{wang2016structural}, or matrix factorization~\cite{ou2016asymmetric,qiu2018network}. Similar to GNN-based representations, a clustering algorithm on the embeddings can be used to detect communities from these representations. Node embeddings can be seen as a generalization of dimensionality reduction methods, and tend to preserve the structure, but disregard node attributes. 

A few recent works address the problem of attributed node embeddings through matrix factorization~\cite{yang2015network} or deep models~\cite{gao2018deep}. \dmw{None of these models are designed for community detection. \dmi{AGC~\cite{zhang2019attributed} proposes a combined score based on spectral clustering on top of a GNN representation.}}
}

\section{\dmi{Communities and Modularity}}\label{sec:main}
Consider an \emph{attributed graph} $\Gr = (\V, \E, \A)$ where $\V = \{v_1, .., v_\nn\}$ is a set of $\nn$ nodes, $\E \subseteq \V \times \V$ is a set of edges and $\A = \{a_1, ..., a_\na\}$ is set of $\na$ attributes. Each node $v_i$ has an associated vector $\xvec_i \in \reals^\na$ of real features for each attribute. 
The node features \d form an ${\nn \times \na}$ matrix $\xmat \in \reals^{\nn \times \na}$ where each node-feature vector $\xvec_i$ is a \dmw{row in such matrix.} 
The \emph{adjacency matrix} is a matrix representation  $\amat$ of the graph's structure, where $\amat_{ij} = 1$ if $(v_i,v_j) \in \E$, and $0$ otherwise. The \emph{degree} $\dg_i$ of a node $v_i$ is the number of neighbors of node $i$, i.e., $\dg_i = \sum_{j=0}^\nn \amat_{ij}$; $\dgvec$ is the vector containing the degree $\dgvec_i=\dg_i$ of all nodes, and $\dgmat$ is the diagonal degree matrix.

\begin{problem*}[Attributed graph community detection.] \noindent We aim to assign each node to at least one of $k$ communities, $\Cl_1, ..., \Cl_k$, such that \dmk{a score of \emph{community cohesiveness} is maximized.} The cluster assignment is a probability vector $\cvec_i$ indicating the probability of node $v_i$ belonging to community $\Cl_j$. \dmw{Cluster assignments form a matrix $\cmat \in [0,1]^{\nn \times k}$ where row $i$ contains node $i$'s cluster assignment $\cvec_i$.}
\end{problem*}

\smallskip
\ia{One of the determinant choices for community detection algorithms is the definition of the community cohesiveness score that determines the quality of the cluster assignments. \rev{We now} review modularity~\cite{newman2006modularity}, a popular measure for community detection.}


\subsection{Modularity}\label{ssec:modularity}

\emph{Modularity}~\cite{newman2006modularity} $Q(\Gr;\Cl)$ measures the quality of a partition $\Cl$ of the nodes of the graph $\Gr$; a high modularity score indicates that node grouped by $\Cl$ have dense internal connections and sparse connections to outside nodes. More specifically, modularity captures the difference in density between the edges inside a community $\cvec_i$ and the edges of a fixed null model:

\begin{equation}\label{eq:modularity1}
    Q(\Gr;\Cl)=\frac{1}{4\mid\E\mid}\sum_{s=1}^k\sum_{ij}\left(\amat_{ij}-\frac{{\dg_i}{\dg_j}}{2\mid\E\mid }\right)\cmat_{is}\cmat_{js}.
\end{equation}

\noindent The quantity $\frac{{\dg_i}{\dg_j}}{2\mid\E\mid }$ is the null-model representing the probability that two nodes $v_i, v_j$ are connected by chance. 
The null model in the modularity score is the rewiring model, in which each node $v_i$ preserves its degree $\dg_i$ but connects randomly to any other node in the graph.
By defining the \emph{modularity matrix} $\bmat$ as $\bmat_{ij} = \amat_{ij}-\frac{{\dg_i}{\dg_j}}{2\mid\E\mid}$  Eq.~\ref{eq:modularity1} simplifies into
\begin{equation}\label{eq:mod_matrix}
    Q(\Gr;\Cl)=\frac{1}{4\mid\E\mid}\tr(\cmat^\top\bmat\cmat)
\end{equation}




\vspace{-.2cm}
\dmw{
\para{Limits and pitfalls.}  \noindent Modularity maximization is one of the most popular methods for community detection~\cite{fortunato2010community}. However, its \emph{direct} maximization may fail to provide optimal communities. \dmk{As shown in \cite{fortunato2007resolution}, modularity may fail to recognize \dmi{communities} that fall below a graph-specific size}. Furthermore, modularity is a measure for discrete partitioning and \rev{does not perform well in the case of overlapping communities~\cite{devi2016analysis}}. \dmi{In the following section, we show how to overcome these limitations of modularity by combining the expressiveness of Graph Neural Networks with a novel contrastive modularity loss that captures both overlapping and non-overlapping communities.} 
}

\section{Our solution: \ourmethod}\label{sec:solution}
\dmi{The modularity objective in Eq.~\ref{eq:mod_matrix} is \NPhard, but can be solved efficiently with a spectral approach similar to spectral clustering~\cite{newman2006modularity} if we allow matrix $\cmat$ to be real rather than binary. This relaxed objective admits as solutions the $k$ leading eigenvalues of the matrix $\bmat$. \dmw{This convenient relaxation enables soft clustering assignments and, in principle, overlapping community detection. 

To circumvent the modularity's resolution limit and capture interactions among nodes that are not directly connected, we further assume that $\cmat$ is the output of a Graph Neural Network model.} 

\mpara{Graph Neural Network approach.} \noindent Graph Neural Networks (GNNs)~\cite{kipf2017GCN, velivckovic2018deep, hamilton2017inductive, bronstein2017geometric, battaglia2018relational} transform the node attributes by nonlinear aggregation of attributes of each node's neighbors. By virtue of this aggregation mechanism, these networks are called message passing~\cite{battaglia2018relational}. We now review the \dmw{Graph Convolutional Network (GCN) model~\cite{kipf2017GCN}.} 
We denote as $\xmat^{[0]}$ the initial node attributes $\xmat^{[0]} = \xmat$,  
\[\hat{\amat} = \dgmat^{-1/2}(\adj + \eye)\dgmat^{-1/2}\]  
the normalized adjacency matrix with self-loops, and $\wmat^{[t]}$ the weight matrix at layer $t$, which encodes the parameters of the network. 
The $t{+}1$ layer $\xmat^{[t+1]}$ is
\begin{equation*}
    \xmat^{[t+1]} = \sigma(\hat{\amat}\xmat^{[t]}\wmat^{[t]})
\end{equation*}
The function $\sigma$ is a non-linear activation function, such as softmax, SeLU, or ReLU. The matrix $\wmat^{[0]}$ is randomly initialized, typically as $\wmat^{[0]}{\sim}\mathcal{N}(0,1)$. The parameters $\wmat$ are learned via stochastic gradient descent on a supervised or unsupervised loss function. 
The result of a GNN in the last layer $T$ is a matrix $\xmat^{[T]}$ which rows are embeddings of a node in a $d$-dimensional space. 

To train a GNN, we need to specify a differentiable loss function. For instance, in the node classification task, the loss function is typically the binary cross-entropy. An optimizer, such as ADAM~\cite{kingma2014adam}, finds the parameters $\wmat^{[1]}, ..., \wmat^{[T]}$ that minimize the loss function. 

The choice of the architecture and the loss function are determinant choices for GNNs. In what follows, we present our model \ourmethod that integrates the simplicity of single-objective community detection with the power of combined scores, by virtue of} a new loss function that encourages \dmk{robust} community memberships while maintaining consistent separation between dissimilar nodes.
\subsection{\ourmethod Loss function}
\label{ssec:loss}
\dmw{
We build our loss function based on community modularity (Eq.~\ref{eq:mod_matrix}). 
We start by showing that the entire matrix $\cmat^\top\bmat\cmat$ can be interpreted as the modularity \dmk{\emph{across}} communities. 
Afterward, we introduce our contrastive loss and show how such a loss aims to detect overlapping and non-overlapping communities alike.}

\subsubsection{$\cmat^\top\bmat\cmat$ as modularity across communities.} 
\dmk{We observe that} $\cmat^\top\bmat\cmat$ encodes the modularity matrix at the community scale
\begin{align*}
    \cmat^\top\bmat\cmat &= \cmat^\top\left(\adj - \frac{\dvec\dvec^\top}{2\mid\E\mid}\right)\cmat = \underbrace{\cmat^\top\adj\cmat}_{\adj^\Cl} - \frac{1}{2\mid\E\mid}\underbrace{(\cmat^\top\dvec)(\dvec^\top\cmat)}_{\dmat_\Cl\dmat_\Cl^\top} =  Q_M
\end{align*}
We refer to $Q_M$ as the community-wise modularity matrix.

In the simple setting where $\cmat$ is binary, such that $\cmat \in \{0, 1\}^{\nn \times k}$, then $Q_M$ reasonably represents the modularity across the community graph. Note that $\adj^\Cl$ is \dmk{the weighted adjacency matrix of a graph where nodes are communities and the weight $\adj^\Cl_{ij}$ is twice} the number of edges between community $\Cl_i$ and community $\Cl_j$. The diagonal entries $\adj^\Cl_{ii}$, therefore, represent the weight from community $i$ to itself and are equal to double the number of edges between the nodes within community $\Cl_i$. \dmk{We also observe} that $\dmat_\Cl = \cmat^\top\dvec$ is the community degree matrix and that $\dmat_\Cl \dmat_\Cl^\top / (2\mid\E\mid)$ represents the likelihood of an edge existing between communities. 
As such, we can interpret $Q_M$ as the modularity of the graph in \dmk{which nodes} are replaced with their corresponding communities.

In the more practical case of non-binary community memberships with $\cmat \in [0,1]^{\nn \times k}$, we can interpret $Q_M$ as the modularity across ``fuzzy'' communities, where each entry of the matrix is proportional to the corresponding community membership strengths.

\rev{We can now state our objective as maximizing the diagonal values of $Q_M$ while minimizing off-diagonal entries that correspond to dissimilar communities. Clearly, then, our target diagonal values should be 1. However, setting the target off-diagonal values to 0 would penalize overlapping community detection. For this reason, we define a target distribution $y \in \mathbb{R}^{2k}$ as follows:
\begin{equation}
    y_i = \begin{cases}
        1 & \text{ if } i \leq k \\ 
        \delta & \text{ otherwise}
    \end{cases}
\label{eq:target_function}
\end{equation}
where $\delta$ is a threshold parameter set to $0$ in the non-overlapping setting and a pre-determined value in the overlapping setting\footnote{$\delta=0.85$ in all datasets in our experimental cohort.}. \revm{A $2k$ vector is necessary to enforce the similarity between the first $1,..,k$ elements and dissimilarity among the next $k+1, ..., 2k$ elements.} 
Under the distribution in \revm{Eq.}~\ref{eq:target_function}, we optimize for community-wide modularity by matching intra-community similarities $(Q_M)_{ii}$ to the target $y_{i; i \leq k}$ and inter-community similarities $(Q_M)_{jl; l \neq j}$ to the target $y_{j; j>k}$. Thus, our loss function becomes}

\rev{\vspace{-0.2cm}
\begin{equation} 
\mathcal{L}_{\ourmethod}=-\frac{1}{2k}\sum_{i=1}^{k} (\underbrace{( \yvec_i \log (\Diag(\sigma(Q_M))_i)}_{\text{intra-community}} + \underbrace{(1 - \yvec_{k+i})\log (1- \Diag(\mathcal{P}(\sigma(Q_M)))_i }_{\text{inter-community}})
\label{eq:objective}
\end{equation}}

where $\Diag$ \dmk{extracts the vector of the diagonal of a matrix}, $\mathcal{P}$ returns a random row-permuted matrix, and $\sigma$ is the element-wise sigmoid. The row permutation ensures that every community is compared repulsively to another community, as the post-permutation diagonal contains the community modularity between separate clusters.  \revm{Although the loss allows for including multiple permutations $\mathcal{P}$ of the modularity matrix, in practice, we only consider one as we find that this choice strikes a balance between speed and quality.} Thus, this loss function has the straightforward interpretation of clustering similar groups of nodes while encouraging separation between dissimilar ones.

Note that $\mathcal{L}_{\ourmethod}$ has a natural relationship to cross-entropy and contrastive objective functions. \rev{In the non-overlapping setting, it corresponds to the cross-entropy loss as it represents the KL divergence between Bernoulli random variables. Our target is not a probability distribution in the overlapping setting, however, requiring us to scale the loss by $(1+\delta)$ to recover the cross-entropy interpretation.}

\subsubsection{A loss for overlapping and non-overlapping communities}
\label{sssec:theory}

\captionsetup[wrapfigure]{justification=centering,labelfont={bf},labelformat={default},labelsep=period,name={\;\;\revm{Figure}}}
\begin{wrapfigure}{r}{.15\textwidth}
\vspace{-8mm}
\usetikzlibrary{positioning, fit, shapes.geometric}
    \begin{tikzpicture}[
      mycircle/.style={
         circle,
         draw=black,
         fill opacity = 0.3,
         text opacity=1,
         inner sep=0pt,
         minimum size=14pt,
         font=\small},
      node distance=0.1cm and 0.3cm
      ]
      \node[mycircle] (v3) {$v_3$};
      \node[mycircle, above left=of v3] (v1) {$v_1$};
      \node[mycircle,below left=of v3] (v2) {$v_2$};
      \node[mycircle,above right=of v3] (v4) {$v_4$};
      \node[mycircle,below right=of v3] (v5) {$v_5$};

    \node[ellipse, draw=cycle1, dashed, inner sep = -0.5mm, label={[cycle1]above:$c_1$}, fit=(v1) (v2) (v3)] (c1) {};
    \node[ellipse, draw=cycle2, dashed, inner sep = -0.5mm, label={[cycle2]above:$c_2$}, fit=(v3) (v4) (v5)] (c2) {};
    \foreach \i/\j in {
        v1/v2,
        v1/v3, 
        v2/v3, 
        v3/v4,
        v3/v5,
        v4/v5}
       \draw (\i) -- (\j);
    \end{tikzpicture}
    \centering
    \caption{}
    \vspace{-6mm}
    \label{fig:bowtie_graph}
\end{wrapfigure}

\rev{Our loss in Equation~\ref{eq:objective} clearly encourages non-overlapping community structure by maximizing the diagonal of $Q_M$ and minimizing the off-diagonal. It is less clear whether such a loss also supports overlapping community detection. To this end, we consider the bowtie graph \revm{depicted in Figure~\ref{fig:bowtie_graph}} with $5$ vertices and edges $\mathcal{E} = \{[v_1, v_2], [v_1, v_3], [v_2, v_3], [v_3, v_4], [v_3, v_5], [v_4, v_5]\}$; $v_{1, 2, 4, 5}$ have degree $2$, $v_3$ has degree 4. The optimal overlapping clustering then groups vertices $v_1, v_2, v_3$ into community $c_1$, $v_3, v_4, v_5$ into $c_2$ with $v_3$ shared among $c_1$ and $c_2$.

If we assume our loss is minimized by non-overlapping communities, it would incentivize orthogonal binary community indicator vectors. WLOG, let $c_1^{n} = [1, 1, 0, 0, 0]^{\top}$ and $c_2^{n} = [0, 0, 1, 1, 1]^{\top}$ be two such non-overlapping communities. Comparing this to the optimal overlapping clustering $c_1^{o} = [1, 1, 0.5, 0, 0]^{\top}$ and $c_2^{o} = [0, 0, 0.5, 1, 1]^{\top}$, we obtain
\[ \mathcal{L}_{\ourmethod}(c_1^{n}, c_2^{n}) = 0.124 > \mathcal{L}_{\ourmethod}(c_1^{o}, c_2^{o}) = 0.094 \]
An exhaustive search over all possible communities shows that the minimum of the loss function is the clustering $\mathcal{C} = [c_1^{o}, c_2^{o}]$.  \revm{As such, the loss already encourages overlapping communities. Yet, the value of $\delta$ can increase to allow for additional overlap-sensitivity if necessary. In the future, one could consider varying $\delta$ on a per-community basis.}

We support the above example with an ablation study across datasets. Table~\ref{tbl:ablation} shows that optimizing both elements of the contrastive loss yields the best overlapping \emph{and} non-overlapping NMI.
}

\begin{table}[htb]
\centering
\setlength{\tabcolsep}{3pt}
\newcolumntype{C}{>{\centering\arraybackslash}X}
\newcolumntype{R}{>{\raggedleft\arraybackslash}X}
\begin{tabularx}{\textwidth}{l*{7}{R}}
& \multicolumn{3}{c}{\textit{non-overlapping}} & \multicolumn{4}{c}{\textit{overlapping}} \\
\cmidrule(lr){2-4} \cmidrule(lr){5-8}
 & \multicolumn{1}{C}{Cora} & \multicolumn{1}{C}{Citeseer} & \multicolumn{1}{C}{Pubmed}  & \multicolumn{1}{C}{fb\_348} & \multicolumn{1}{C}{fb\_414} & \multicolumn{1}{C}{fb\_686} & \multicolumn{1}{C}{fb\_1684} \\ 
\midrule
Intra-community  & {51.8} & {28.0} & {20.9} & {23.3} & {33.5} & {14.6} & {25.4} \\
Inter-community  & {0.0} & {0.0} & {0.0} & {16.3} & {23.7} & {11.5} & {25.7} \\  
\rowcolor{cycle2!20}\ourmethod & \textbf{57.4} & \textbf{41.0} & \textbf{25.0} & \textbf{33.9} & \textbf{59.9} & \textbf{22.1} & \textbf{33.3}
\end{tabularx}
\vspace{6pt}
\caption{\rev{NMI scores optimizing only \emph{intra-community} similarity with target function $y_{i; i < k}$,  \emph{inter-community} similarity with target function $y_{i; i > k}$, and the \ourmethod objective in  Equation~\ref{eq:objective}.}}
\label{tbl:ablation}
\end{table} 

\subsection{\ourmethod architecture} 
\label{ssec:architecture}
\dmw{The main purpose of our GCN is to learn the community assignment matrix $\cmat$ using the graph structure and the node attributes.
Our architecture is a two-layer GCN~\cite{kipf2017GCN}: }
\begin{equation}
\begin{aligned}
GCN(\hat{\adj},\xmat) &= \texttt{\rev{RReLU}}(\hat{\adj}\cdot\underbrace{\texttt{\rev{SiLU}}(\hat{\adj}\xmat\wmat^{[0]})}_{\xmat^{[1]}}\wmat^{[1]})\\
\rev{\text{~~where~~}&\hat{\amat} = \dgmat^{-1/2}(\adj + \eye)\dgmat^{-1/2}}
\end{aligned}
\label{eq:model}
\end{equation}
\vspace{-.2cm}

\dmw{The last layer of our GCN outputs community \rev{assignments via}} 
\begin{equation*}
\rev{\cmat = GCN(\hat{\adj}, \xmat)}
\end{equation*}
\dmw{This architecture, although simple, allows for propagating information over the entire graph, thus capturing relationships within the graph's structure and the nodes' attributes.}

\section{Experiments}\label{sec:experiments}


\dmw{In this section, we empirically evaluate \ourmethod in comparison with state-of-the-art approaches for community detection on several benchmark graph datasets. We analyze our results in both non-overlapping community detection (graph partitioning), and in overlapping community detection in Section~\ref{ssec:nonoverlapping} where nodes may be assigned to more than one community (as discussed in Section~\ref{ssec:overlapping}). \dme{We further analyse the stability of the performance~\ref{ssec:stability} and sensitivity of our approach to its few hyperparameters (Section~\ref{ssec:tuning}).} 
}

\dme{We implement \ourmethod using PyTorch version 1.10.0 and Python v3.8. We release the implementation of \ourmethod at \url{https://github.com/AU-DIS/UCODE}. We evaluate our methods on a 14-core Intel Core i9 10940X 3.3GHz machine with 256GB RAM.}

\mpara{Our method. }
\ourmethod outputs an assignment matrix $\cmat$  \dme{where \rev{ $c_{ij}$} represents the likelihood} of node $v_i$ belonging to community $j$. For non-overlapping community detection, we assign the node to the community with the highest \rev{score, i.e., $\argmax_j c_{ij}$}.

In additional experiments, we also investigate a second version, \textbf{\ourmethodk}, which \dmk{applies the \kmeans algorithm on the representations obtained by the \rev{RReLU function}} in Eq.~\ref{eq:model}. Studying this version, we show the benefit of our method compared to decoupled community detection approaches. The results suggest that \kmeans contributes only marginal quality improvement, which confirms the validity of our efficient end-to-end loss function for community detection. 

\mpara{Adapting to regularization.} We note that the values in $Q_M$ can be positive or negative and are not necessarily bounded. The sigmoid is thus necessary in order to calculate the cross-entropy to the target distribution. However, we found empirically that the division by $4{\mid}\mathcal{E}{\mid}$ in Eq. \ref{eq:mod_matrix} settles the values in $\cmat^\top\bmat\cmat$ close to 0, leaving the sigmoid outputs near $1/2$. 
To this end, we apply a logarithm in $\cmat^\top\bmat\cmat$ that preserves the ordering but amplifies the values. In preliminary experiments, we empirically confirmed that this approach sufficiently amplifies the values so as to achieve good performance when using network regularization.

\mpara{Competitors.} We collect results for a number of state-of-the-art non-overlapping \revm{(Section~\ref{sssec:app:non-ovelapping} in the appendix)} and overlapping \revm{(Section~\ref{sssec:app:overlap} in the appendix)} community detection methods.

\dme{\mpara{Quality measures.} For both tasks of overlapping and non-overlapping community detection, we provide the Normalized Mutual Information (NMI) between the cluster assignments and the ground-truth communities. In addition, for non-overlapping community detection we provide the pairwise F1 score between all node pairs and their corresponding ground-truth community; we also provide two intrinsic quality measures, namely modularity (Eq.~\ref{eq:modularity1}) and network conductance~\cite{Yang2015conductance}.} The network conductance ($\mathcal{C}$)
measures how well-connected the nodes in the communities are related to the escape probabilities of random walks. Modularity ($Q$)~\cite{newman2006modularity} assesses whether intra-community nodes are more densely connected than their inter-community counterparts. We report the average value of each measure over $10$ runs of the algorithms.


\begin{table}[!h]
\newcolumntype{C}{>{\centering\arraybackslash}X}
\newcolumntype{R}{>{\raggedleft\arraybackslash}X}

\begin{tabularx}{\linewidth}{lRRRRRC}
\toprule
\textbf{Dataset} & $\mid\V\mid$ & $\mid\E\mid$ &  $\mid\A\mid$ & Dens. & Comm. & Overl.\\ 
\midrule
\cora   & 2\,700 &  5\,300 & 	 1\,433  & $.04$ 	& 7 & \xmark\\
\citse & 3\,300 & 4\,600  & 	 3\,703  & $.04$ 	& 6 & \xmark \\ 
\pubmed  & 19\,700 & 44\,300  &  500     & $.01$ 	& 3 & \xmark\\          
\amzph & 7\,700 & 71\,800 & 	 745     & $.11$ 	& 8 & \xmark\\
\amzpc & 13\,700 & 143\,600 & 	 767     & $.07$ 	& 10 & \xmark\\
\coacs & 18\,300 & 81\,900 & 	 6\,805  & $.01$ 	& 15 & \xmark\\
\coaphy & 34\,500 & 247\,900 & 	 8\,415  & $.01$ 	& 5 & \xmark\\
\fb-348  & 224 &  3\,200 & 	 21      & $12.80$ 	&  14 & \cmark\\
\fb-414 & 150 &  1\,700  & 	 16      & $15.10$ 	& 7 & \cmark\\ 
\fb-686  & 168 & 1\,600  & 	 9       & $11.30$ 	& 14 & \cmark\\ 
\fb-698 & 61 & 270 & 		 6		 & $14.80$ 	& 13      & \cmark\\
\fb-1684 & 786 & 14\,000 & 	 15      & $4.50	$& 17 & \cmark\\
\fb-1912  & 747 & 30\,000 & 	 29      & $10.80$	& 46 & \cmark\\    
\eng & 14\,900 & 49\,300 & 		 4\,800  & $.04		$& 16 & \cmark\\  
\bottomrule
\end{tabularx}
\caption{Datasets and their main characteristics.}
\label{tbl:datasets}
\end{table}

\mpara{Data.} We perform experiments on $14$ real-world graphs with non-overlapping and overlapping communities. The largest graph has $34.5K$ nodes and $247K$ edges. Further details on the datasets, quality measures and parameter settings can be found in Table~\ref{tbl:datasets}. \dme{Our choice of datasets includes graphs with different types of communities, density and attributes, as well as the largest networks evaluated by the competitors.}

\begin{itemize}[leftmargin=*, noitemsep, topsep=0pt]
\item 
\textbf{\cora}, \textbf{\citse}, and \textbf{\pubmed}~\cite{sen2008collective} are co-citation networks among papers where attributes are bag-of-words representations of the paper's abstracts, and labels are paper topics. 
\item 
\dmk{\textbf{\amzph} and \textbf{\amzpc}~\cite{shchur2018pitfalls}  are subsets of the Amazon co-purchase graph with the frequency of products purchased together; attributes are bag-of-words representations of product reviews, and class labels are product categories.
\item 
\textbf{\coacs} and \textbf{\coaphy}~\cite{shchur2018pitfalls} are co-authorship networks based on the MS Academic
Graph (MAG) for the computer science and physics fields respectively;  attributes are collections of paper keywords; class labels indicate common fields of study.} 
\item\dmi{\textbf{Fb-X} datasets~\cite{mcauley2014discovering} are ego-nets from Facebook where X is the id of the central node. 
\item\textbf{Eng}~\cite{shchur2019overlapping} is a co-authorship graph from MAG.} 
\end{itemize}


\subsection{Non-Overlapping Community Detection}\label{ssec:nonoverlapping}

We begin our experimental evaluation with an overall comparison of methods for non-overlapping community detection across different datasets. \revm{We compare with the methods described in Section~\ref{sssec:app:non-ovelapping} in the appendix}. We additionally include \textbf{\nocd}~\cite{shchur2019overlapping}, a state-of-the-art GNN for overlapping community detection. To obtain non-overlapping clusters, we assign each node to the cluster with the highest probability.

\dme{\mpara{\ourmethod parameter setup.} 
We train \ourmethod for $1000$ epochs, which shows consistent results across datasets and tasks. We use two GCN layers with a hidden dimension $256$. We default to producing $k=16$ communities for all datasets as this choice is consistent 
with \mincut~\cite{bianchi2020spectral} and \dmon~\cite{tsitsulin2020graph} and, in a set of preliminary experiments, we found the performance with $k = 8$ and $k=32$ to give inferior results.
We apply batch normalization in both internal layers \dmk{and set a learning rate $10^{-3}$ for the Adam optimizer~\cite{kingma2014adam} for learning.} 
\rev{We add weight decay to both weight matrices with regularization strength $\lambda = 10^{-1}$.}

We additionally experimented with GraphSAGE~\cite{hamilton2017inductive} for the internal propagation layer, but opt for GCN~\cite{kipf2017GCN} due to the superior performance in our analyses. }


\subsubsection{Analysis of ground-truth communities}




\begin{figure*}[htb]
\centering
\pgfplotsset{/pgfplots/error bars/error bar style={draw=gray, color=gray, thick}}

\begin{subfigure}[b]{\textwidth}
\pgfplotstableread{results/nmi-nonoverlapping-CI.csv}\datatable
\begin{tikzpicture}
\begin{axis}[
    ybar,
    ylabel={NMI},
    ymin=0,
    ymax=100,
    width=\textwidth,
    bar width=0.7ex,
    align=right, 
    height=3.8cm,
    xticklabels=\empty,
    legend columns=7,
    legend style={at={(-0.1,1.35)},anchor=north west, font=\small},
    legend entries={kmeans, Louvain, MinCut, DGI+kmeans, NOCD, DMoN, \ourmethod},
    xtick pos=bottom,
    ytick pos=left,
]
\addplot [cycle6, fill=cycle6!20, error bars/.cd, y dir=both, y explicit] 
table [x expr=\coordindex, y={kmeans}, y error={Kmeans-CI}]{\datatable};

\addplot [cycle7, fill=cycle7!20, error bars/.cd, y dir=both, y explicit] 
table [x expr=\coordindex, y={Louvian}, y error={Louvian-CI}]{\datatable};

\addplot [cycle2, fill=cycle2!20, error bars/.cd, y dir=both, y explicit] 
table [x expr=\coordindex, y={MinCut}, y error={MinCut-CI}]{\datatable};

\addplot [cycle3, fill=cycle3!20, error bars/.cd, y dir=both, y explicit] 
table [x expr=\coordindex, y={DGI+kmeans}, y error={DGI+kmeans-CI}]{\datatable};

\addplot [cycle4, fill=cycle4!20, error bars/.cd, y dir=both, y explicit] 
table [x expr=\coordindex, y={NOCD-X}, y error={NOCD-X-CI}]{\datatable};

\addplot [cycle5, fill=cycle5!20, error bars/.cd, y dir=both, y explicit] 
table [x expr=\coordindex, y={DMoN}, y error={DMON-CI}]{\datatable};

\addplot [cycle1, fill=cycle7!20, pattern color=cycle1!70, pattern=crosshatch, error bars/.cd, y dir=both, y explicit] 
table [x expr=\coordindex, y={UCoDe}, y error={UCoDe-CI}]{\datatable};

\end{axis}%
\end{tikzpicture}
\vspace*{-3mm}
\end{subfigure}
\begin{subfigure}[b]{\textwidth}
\centering
\pgfplotstableread{results/f1-nonoverlapping-CI.csv}\datatable
\begin{tikzpicture}
\begin{axis}[
    ybar,
    ylabel={F1},
    ymin=0,
    ymax=100,
    width=\textwidth,
    align=right,
    bar width=0.7ex, 
    height=3.8cm,
    xtick=data,
    xticklabels from table={\datatable}{Data},
    y label style={at={(-0.05, 0.5)}},
    xtick pos=bottom,
    ytick pos=left,
]

\addplot [cycle6, fill=cycle6!20, error bars/.cd, y dir=both, y explicit] 
table [x expr=\coordindex, y={kmeans}, y error={kmeans-CI}]{\datatable};

\addplot [cycle7, fill=cycle7!20, error bars/.cd, y dir=both, y explicit] 
table [x expr=\coordindex, y={Louvain}, y error={Louvain-CI}]{\datatable};

\addplot [cycle2, fill=cycle2!20, error bars/.cd, y dir=both, y explicit] 
table [x expr=\coordindex, y={MinCut}, y error={MinCut-CI}]{\datatable};

\addplot [cycle3, fill=cycle3!20, error bars/.cd, y dir=both, y explicit] 
table [x expr=\coordindex, y={DGI+kmeans}, y error={DGI+kmeans-CI}]{\datatable};

\addplot [cycle4, fill=cycle4!20, error bars/.cd, y dir=both, y explicit] 
table [x expr=\coordindex, y={NOCD-X}, y error={NOCD-X-CI}]{\datatable};

\addplot [cycle5, fill=cycle5!20, error bars/.cd, y dir=both, y explicit] 
table [x expr=\coordindex, y={DMoN}, y error={DMoN-CI}]{\datatable};

\addplot [cycle1, fill=cycle1!20, pattern color=cycle1!70, pattern=crosshatch, error bars/.cd, y dir=both, y explicit] 
table [x expr=\coordindex, y={UCoDe}, y error={UCoDe-CI}]{\datatable};


\end{axis}%
\end{tikzpicture}
\end{subfigure}
\caption{NMI, \revm{F1, and confidence intervals}, for non-overlapping community detection.}
\label{fig:f1-nmi-nonoverlapping}
\end{figure*}
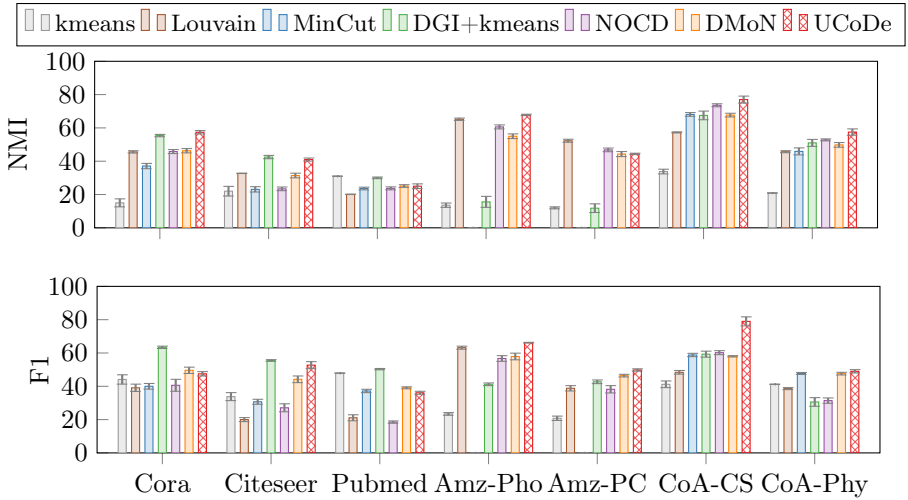

\dme{We compare the methods in terms of NMI and F1-score with respect to ground-truth communities.} As \dme{Figure~\ref{fig:f1-nmi-nonoverlapping}} confirms, \ourmethod is the most robust choice for non-overlapping communities across datasets. Regardless of dataset characteristics, we observe that \ourmethod attains competitive results even where existing approaches under-perform \dme{in several datasets}. Indeed, a more detailed analysis reveals that \ourmethod ranks \textit{on average} higher than any other competitor \revm{(Section~\ref{sssec:app:ttest} in appendix)}. \dme{The additional \kmeans clustering offered to \dgi and \ourmethodk offers a competitive edge only on three of the seven datasets.} Further, note that on the denser \amzpc and \amzph, methods like \mincut and \dcrn fail to converge. They provide overall lower scores, indicating that graph pooling and combined-objectives are not viable approaches for the community detection task. Our method outperforms traditional methods, such as \kmeans, demonstrating an advantage of a graph-learning approach over attribute clustering to capture the structural characteristics of a graph. 
\nocd fares relatively good against methods explicitly targeting non-overlapping communities, but still fails to provide competitive results against \ourmethod.  

In conclusion, there is no clear second choice, promoting \ourmethod to be the method of choice, as it shows consistent behavior across datasets.


\subsubsection{Analysis of conductance and modularity}

\dme{We now turn our attention to intrinsic measures to analyze the impact of the various objective functions on community connectedness. Table~\ref{tbl:C-Q-nonoverlapping} reports conductance ($\mathcal{C}$) and modularity ($Q$).} \ourmethod shows the best performance in terms of conductance, which means that \ourmethod is particularly good at identifying well-connected communities. This makes sense, as our loss function specifically encourages high intra-connections and low inter-connections. 

At the same time, \dmon, which optimizes for modularity, does not consistently attain the best modularity. 
\dme{Yet, \ourmethod attains modularity superior to \dmon in most datasets, although not explicitly encouraging modularity. This indicates that the contrastive loss in \ourmethod indeed yields a more nuanced community structure than can be obtained through optimizing modularity alone. This is even more notable when considering the other measures where \ourmethod outperforms \dmon.}


\dmk{In conclusion, the empirical evaluation clearly shows that our model is highly robust and widely applicable in the non-overlapping setting, obtaining competitive results across the evaluation metrics and datasets rather than targeting any single one. We note that methods that directly optimize modularity achieve good modularity scores at the expense of performance on other measures. \ourmethod instead achieves competitive results across every metric with little-to-no hyperparameter tuning.}

\begin{table*}[th]
\footnotesize

\setlength{\tabcolsep}{3pt}
\newcolumntype{C}{>{\centering\arraybackslash}X}
\newcolumntype{R}{>{\raggedleft\arraybackslash}X}

\begin{tabular}{l*{15}{r}}
\toprule
 & \multicolumn{2}{c}{\cora} & \multicolumn{2}{c}{\citse} & \multicolumn{2}{c}{\pubmed} &
 \multicolumn{2}{c}{\amzph}&
  \multicolumn{2}{c}{\amzpc} &
   \multicolumn{2}{c}{\coacs}&
  \multicolumn{2}{c}{\coaphy}\\
 \cmidrule(lr){2-3} \cmidrule(lr){4-5} \cmidrule(lr){6-7} \cmidrule(lr){8-9}
  \cmidrule(lr){10-11} \cmidrule(lr){12-13} \cmidrule(lr){14-15} 
 \textit{method} & $\mathcal{C}$ & $Q$ & $\mathcal{C}$ & $Q$ & $\mathcal{C}$ & $Q$ &$\mathcal{C}$ & $Q$& $\mathcal{C}$ & $Q$ &$\mathcal{C}$ & $Q$ & $\mathcal{C}$ & $Q$  \\  
 \midrule

 \kmeans   & 19.0          & 64.0               & 26.1          & 54.2         & 19.7          & 54.2       & 16.9 & 63.7      &  83.0   & 4.0       &   45.9        &   20.9    &       46.0     & 33.3    \\

DCRN & \uline{11.0} & 71.0 & \uline{5.6} & 76.6 & \textbf{7.8} & 0.0 & - & - & - & - & 21.3 & 70.0 & \textbf{13.1} & 65.3 \\
 
 \dgi         &12.4          & 70.7           & 6.1       & 74.4  & \uline{12.4}          & 52.7          & 49.3 & 22.4  & 72.0          &  12.6 &   33.6   &    58.6 &    38.6     &   51.2 \\
 
 \mincut             & 22.0          & 70.3            & 11.6          & 80.5          & 34.8          & 58.1  &  - & -   & - & - &  \uline{19.8} &       \textbf{72.8} & 28.8   &  62.9 \\

\nocd               & 14.0  & \textbf{78.3}   &  6.5   &  \textbf{84.0} &  22.2  & 64.8 & \uline{14.4} & \uline{68.8} & 25.0       &  \textbf{59.0}          &  20.6        &   71.8      &  24.9 &  \uline{65.0}   \\

 \dmon               & 22.3   & 68.1    & \textbf{4.6}   & 75.3        & 17.0          &\textbf{69.2}  & 19.1 & 65.3 & \uline{19.7}   &   55.8      &  20.0    &\uline{72.3}      & 23.8  &  \textbf{65.8} \\
 
\ourmethodk & 12.3  & 72.1    & 8.1  & 74.8 & 9.7 & 54.0& 26.3 & 53.3 &  44.6        &     30.0          &  23.7     &     66.5   &   \uline{15.8}  &    60.9 \\

\ourmethod    &\textbf{10.9}    & \uline{76.1}  & 7.1   & \uline{80.9}  & 17.8  & \uline{65.4}  & \textbf{9.4} & \textbf{69.4} & \textbf{13.4}  &  \uline{56.0}    & \textbf{13.9}     & 70.9   &  18.7  & 63.1   \\

\midrule
\louvain    & 12.5 & 81.3  & 6.2  & 89.1  & 15.2  & 76.9  & 10.1 & 74.7 & 21.0  &   64.4    & 17.3     & 73.6   &  22.7  & 66.5   \\

\bottomrule
\end{tabular}
\caption{\rev{Graph conductance $\mathcal{C}$ (low is better) and modularity $Q$. Best performer in bold; second best performer underlined.} \dme{Louvain is included for reference since a direct comparison is not possible as it is not possible to set the number of communities.}}
\label{tbl:C-Q-nonoverlapping}
\vspace{-2mm}
\end{table*}

\subsection{Overlapping community detection} 
\label{ssec:overlapping}

\dme{Here, we analyze the performance of \ourmethod on overlapping community detection. The list of competitors is described in Section~\ref{sssec:app:overlap} in the appendix.}


\mpara{\ourmethod parameter setup.} 
\dme{While \ourmethod does not require hyperparameter tuning across datasets, it requires small adaptations across tasks to accommodate for the uncertain nature of overlapping communities.}
To reflect the intrinsic dimensionality of each dataset that grows with the number of nodes~\cite{tsitsulin2019spectral}, we set \textbf{the size of the first layer} 
to $128$ while keeping the output layer's size fixed to the number of communities $k$. 
We apply \textbf{batch normalization} after the first graph convolutional layer. We add \textbf{weight decay} to both weight matrices with regularization strength $\lambda = 10^{-2}$. \dme{The rest of the hyperparameters are the same as in non-overlapping community detection. }

We set the \textbf{diagonal elements} of the permuted matrix $\mathcal{P}(Q_M)$ in Eq.~\ref{eq:objective} to a value $\delta \in [0,1]$ to avoid penalizing intra-cluster connections. 
We find experimentally $\delta = 0.85$ to attain good experimental results \emph{on all datasets}, without the need for further tuning.


\rev{
\mpara{Community assignment.} In the overlapping scenario, we set a threshold $p$ for scores $c_{ij}$ above which a node $i$ is assigned to a community $j$. 
We set a threshold that exhibits good average performance on all the datasets, thereby eschewing per-dataset tuning. Our first threshold $p_1$ is the average of the $\exp$ of the assignment scores, i.e., $\frac{1}{nk}\sum_{ij} \exp(c_{ij})$, where the $\exp$ encourages sparsity by distributing the values on the range $[0, +\infty)$.} \rev{ We note in Figure~\ref{fig:nmi-epochs-threshold} that this choice corresponds to elbow points in a grid search.} \rev{For the \nocd model, we set $p_{2}=0.5$ as in their experiments. We evaluate the \dmon model using  $p_{1}$ and $p_{2}$, and $p_{3}= \mathbb{E}[\cmat]$ and  report results with $p_3$ as they were the highest in all experiments.}


\subsubsection{Analysis of ground-truth communities}
\dmw{Overlapping community detection results are given in Table~\ref{tbl:overlapping-gcn1} and verify that
\rev{\ourmethod outperforms the state-of-the-art methods on the majority of datasets. The direct optimization of modularity in \dmon cannot easily detect overlapping communities, as opposed to our contrastive modularity loss. 
More importantly, \ourmethod outperforms \nocd in many cases, a GCN that directly aims to detect overlapping communities. 
Lastly, we note that none of the other methods in Table~\ref{tbl:overlapping-gcn1} obtain comparable results to our method. This suggests that our method, that requires \emph{no hyperparameter tuning}, is an effective choice for \emph{both} overlapping and non-overlapping community detection.}}

\begin{table}[!ht]
\small
\setlength{\tabcolsep}{3pt}
\newcolumntype{C}{>{\centering\arraybackslash}X}
\newcolumntype{H}{>{\setbox0=\hbox\bgroup}c<{\egroup}@{}}
\newcolumntype{R}{>{\raggedleft\arraybackslash}X}
\begin{tabularx}{\columnwidth}{p{1.4cm}*{7}{R}}
\toprule
 \textit{dataset} & \multicolumn{1}{c}{\revc CDE} & \multicolumn{1}{c}{\revc SNMF} & \multicolumn{1}{c}{\revc BigClam} & \multicolumn{1}{c}{\nocd}  & \multicolumn{1}{c}{\revc COPRA} & \multicolumn{1}{c}{\dmon} & \multicolumn{1}{c}{\revc \ourmethod}\\  
 \midrule
Fb-348  & \revc 24.8 & \revc 13.5 & \revc 26.0  & \uline{33.6} & \revc 14.6 & 19.9  & \revc \textbf{33.9} \\
Fb-414 & \revc 28.7 & \revc 32.5 & \revc 48.3 & \uline{53.0} & \revc 45.4 & 39.0  & \revc \textbf{59.5} \\
Fb-686 & \revc 13.5 & \revc 11.6 & \revc 13.8 & \uline{18.5} & \revc 9.0 & 12.6 & \revc \textbf{22.1}  \\
Fb-698  & \revc 31.6 & \revc 28.0 & \revc \textbf{45.6} & 34.4 & \revc \uline{38.5} & 19.8 & \revc 34.9  \\
Fb-1684 & \revc 28.8 & \revc 13.0 & \revc 32.7 & 30.0 & \revc 32.7 & 32.7 & \revc \textbf{33.3}  \\
Fb-1912 & \revc 15.5 & \revc 23.6 & \revc 21.4 & \textbf{35.7} & \revc 27.6 & 24.7 & \revc \uline{33.1}\\
\eng & \revc - & \revc 10.1 & \revc 7.9 & \textbf{33.3} & \revc 1.0 & 28.5  & \revc \textbf{33.2} \\
\bottomrule
\end{tabularx}
\caption{NMI for overlapping community detection; \rev{ CDE, SNMF, and BigClam results are from~\cite{shchur2019overlapping}.}}
\label{tbl:overlapping-gcn1}
\vspace{-4mm}
\end{table}

\dme{\subsection{Stability analysis}
\label{ssec:stability}

After having established the top-performers in the respective tasks, namely \mincut, \dgi,\nocd, and \dmon for non-overlapping community detection, and \nocd for overlapping community detection, we analyze them in terms of variance. Table~\ref{tbl:nmi-variance-nonoverlapping} shows the NMI and the confidence intervals at 95\% level. \ourmethod retains competitive stability across datasets. More interestingly, while our \kmeans variant, \ourmethodk, attains lower variance due to the \kmeans clustering, \ourmethod is typically comparable and sometimes more stable than the competitors. 

In overlapping community detection in Table~\ref{tbl:nmi-variance-overlapping}, we compare only with \nocd that competes with \ourmethod. The probabilistic nature of the two methods is reflected in the deviation, which is typically around $1.0$. However, in most of the cases, the deviation does not affect the final result and shows that \ourmethod is competitive regardless of the variance. 
}


\begin{table*}[!htb]
\vspace{-2mm}
\small
\setlength{\tabcolsep}{5pt}
\newcolumntype{C}{>{\centering\arraybackslash}X}
\newcolumntype{G}{>{\centering\arraybackslash\columncolor{LightGray}}X}
\newcolumntype{H}{>{\setbox0=\hbox\bgroup}c<{\egroup}@{}}
\begin{tabularx}{\textwidth}{lccccccc}
\toprule
Dataset  & \dgi &\mincut & \nocd & \dmon & \ourmethodk & \ourmethod \\ 
\midrule
\cora  & 55.4 $\pm {0.7}$ & 37.1 $\pm {1.6}$ 	& 45.8 $\pm {1.2}$	 & 46.3 $\pm {1.3}$ 	& 55.7 $\pm{0.6}$ & 57.4 $\pm{1.0}$ \\

\citse & 42.5$\pm {1.0}$ & 23.1 $\pm {1.6}$ 	& 23.4 $\pm {1.1}$ 	& 31.4 $\pm {1.4}$ 	&  44.4 $\pm {0.6}$ & 41.0 $\pm {0.9}$ \\ 

\pubmed & 30.0$\pm {0.5}$ & 23.6 $\pm {0.8}$ & 23.7 $\pm {0.9}$ 	& 25.1 $\pm {0.8}$	& 23.5 $\pm{0.7}$ & 25.0 $\pm {1.4}$ \\  

\amzph  & 15.6 $\pm {3.3}$ & - 				& 60.6 $\pm {1.2}$ 	& 55.0 $\pm {1.4}$ 	& 61.0 $\pm{1.0}$ & 67.8 $\pm{0.4}$ \\

\amzpc & 11.8$\pm {2.6}$ & - 				& 46.8 $\pm {1.1}$ 	& 44.3 $\pm {1.5}$ 	& 38.5 $\pm{0.4}$ & {44.4} $\pm {0.4}$ \\

\coacs & 67.5$\pm {2.6}$ & 68.1 $\pm {1.1}$ 	& 73.6 $\pm {0.9}$	& 67.7 $\pm {1.1}$ 	& 78.4 $\pm{0.2}$ & 77.0 $\pm {2.1}$\\

\coaphy & 51.0$\pm {2.1}$ & 45.9 $\pm {2.1}$	& 52.8 $\pm {0.6}$  & 49.8 $\pm {1.4}$	& 55.3 $\pm{1.7}$& 57.5 $\pm{1.9}$ \\

\bottomrule
\end{tabularx}
\caption{\dme{Non-overlapping community detection: NMI and confidence intervals.}}
\label{tbl:nmi-variance-nonoverlapping}
\vspace{-4mm}
\end{table*}


\begin{table}[!ht]
\centering
\small
\setlength{\tabcolsep}{5pt}
\newcolumntype{C}{>{\centering\arraybackslash}X}
\newcolumntype{H}{>{\setbox0=\hbox\bgroup}c<{\egroup}@{}}
\begin{tabular}{p{1.4cm}rr}
\toprule
 \textit{dataset} & \nocd & \ourmethod\\  
 \midrule
Fb-348  & 33.6  $\pm{0.9}$ & 33.9 $\pm {1.1}$\\
Fb-414  & 53.0 $\pm{1.1}$ & 59.5 $\pm{1.1}$ \\
Fb-686 & 18.5 $\pm{1.0}$ & \ 22.1 $\pm{1.2}$ \\
Fb-698  &  34.4  $\pm{0.3}$ & 34.9 $\pm{1.4}$ \\
Fb-1684 & 30.0 $\pm{1.9}$ &  33.3 $\pm{1.2}$ \\
Fb-1912 & 35.7 $\pm{1.6}$ & 33.1 $\pm{0.6}$\\
\eng & 33.3 $\pm{1.8}$ & 33.2 $\pm{0.9}$ \\
\bottomrule
\end{tabular}
\caption{\dme{Overlapping community detection: NMI and confidence intervals.}}
\label{tbl:nmi-variance-overlapping}
\vspace{-4mm}
\end{table}

\subsection{Sensitivity analysis}
\label{ssec:tuning}
\begin{figure*}[!h]
\small
\vspace{-4mm}
\begin{tikzpicture}
\begin{groupplot}[group style={
                      group name=myplot,
                      group size= 3 by 1, horizontal sep=1.1cm},
                      height=3.5cm,
                      width=0.36\linewidth ,
                      title style={at={(0.5,0)},anchor=north,yshift=-11mm},
                      ymin=0,
                      ymax=0.8,
                      every axis plot/.append style={thick},
                      xticklabel style={/pgf/number format/fixed},ymajorgrids]
                      
\nextgroupplot[
	ylabel={\small Loss},
	xlabel={Epochs},
	ymin=0.6,
	ymax=0.7,
	xmin=0, 
	xmax=500,
]
\addplot[color=cycle2] table [x=X, y=Y] {results/cora-loss-ucode.csv};

\nextgroupplot[
	ylabel={\small NMI},
	xlabel={Epochs},
	y label style={at={(-0.14,0.5)}},
	legend entries={DMoN, \ourmethod },
	legend pos=south east,
    xtick pos=bottom,
    ytick pos=left,
	ymin=0,
	ymax=70,
	xmin=0, 
	xmax=500,
]
\addplot[color=cycle1] table [x=X, y=Y, y expr=\thisrowno{1}*100] {results/dmon_fixed.csv};
\addplot[color=cycle2] table [x=X, y=Y, y expr=\thisrowno{1}*100] {results/nmi-cora-epochs-ucode.csv};

\nextgroupplot[
	ylabel={\small Modularity ($Q$)},
	xlabel={Epochs},
	y label style={at={(-0.14,0.5)}},
	legend entries={ DMoN, \ourmethod },
	legend pos=south east,
    xtick pos=bottom,
    ytick pos=left,
	ymin=0,
	ymax=100,
	xmin=0, 
	xmax=500,
]
\addplot[color=cycle1] table [x=X, y=Y, y expr=\thisrowno{1}*100] {results/modularity-cora-dmon-epochs.csv};
\addplot[color=cycle2] table [x=X, y=Y, y expr=\thisrowno{1}*100] {results/modularity-cora-ucode-epochs.csv};
\end{groupplot}

\end{tikzpicture}
\caption{\dml{Training (Cora): \rev{ \ourmethod quickly minimizes the loss (left); NMI increases steadily and achieves $21\%$ higher value than DMoN (center); \ourmethod gradually outperforms DMoN's modularity (right). The alternating convergence pattern is characteristic of contrastive loss.}}}
\vspace{-3mm}
\label{fig:tuning}
\end{figure*}
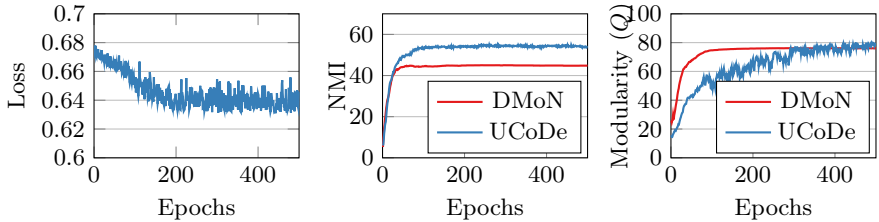

\dm{
We analyse \ourmethod as a function of the number of training epochs. 
\dmw{Figure~\ref{fig:tuning} gives results for non-overlapping community detection for the Cora dataset (other datasets show similar trends).} As expected in \amw{Figure~\ref{fig:tuning} (left)}, the loss decreases with more epochs and converges after only about 100 epochs.
This behaviour is confirmed in the NMI score (center).}
\dmw{Figure~\ref{fig:tuning} \amw{(right)} compares the modularity score for DMoN and \ourmethod\ \dmk{per training epoch.} \rev{We note that while the contrastive loss stabilizes after 200 epochs, the modularity continues to increase until it outperforms DMoN, confirming our analysis in Section~\ref{ssec:loss}.}}

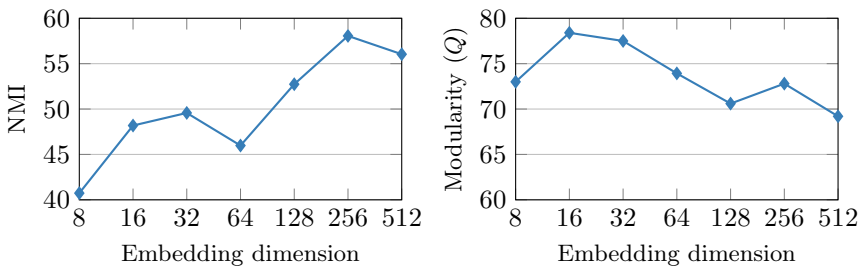
\begin{figure}[!h]
\begin{tikzpicture}
\centering
\begin{groupplot}[group style={
                      group name=myplot,
                      group size= 2 by 1, horizontal sep=1.5cm},
                      height=4.0cm,
                      width=.49\textwidth,
                      title style={at={(0.5,0)},anchor=north,yshift=-11mm},
                      ymin=0,
                      ymax=0.8,
                      every axis plot/.append style={thick},
                      xticklabel style={/pgf/number format/fixed},ymajorgrids]
\nextgroupplot[
	xlabel={\small Embedding dimension},
	ylabel={\small NMI},
	ymin=40,
	ymax=60,
	xmin=8, 
	xmax=512,
	xmode=log,
	xtick={8,16,32,64,128,256,512},
	log ticks with fixed point,
]
\addplot[color=cycle2,mark=diamond*, mark size=2pt] table [x=Dimension, y=NMI, y expr=\thisrowno{1}*100] {results/nmi-dim.csv};
\nextgroupplot[
	xlabel={\small Embedding dimension},
	ylabel={\small Modularity ($Q$)},
	ymin=60,
	ymax=80,
	xmin=8, 
	xmax=512,
	xmode=log,
	y label style={at={(-0.12,0.5)}},
	xtick={8,16,32,64,128,256,512},
	log ticks with fixed point,
]
\addplot[color=cycle2,mark=diamond*, mark size=2pt] table [x=Dimension, y=Modularity] {results/mod-dim.csv};
\end{groupplot}
\end{tikzpicture}
\caption{Impact of the embedding dimension for non-overlapping dataset Cora (similar for other data). The maximum modularity (right) does not correspond to the best NMI (left). The optimal embedding dimension for the intermediate layer is $256$.}
\label{fig:embedding-epochs-threshold}
\end{figure}

\dmi{
We study the impact of the embedding dimension on the quality of the communities. In \amw{Figure~\ref{fig:embedding-epochs-threshold}}, we report both NMI and modularity for the Cora dataset (other datasets show similar trends). For NMI, we note that increasing the dimension is beneficial until the dimension reaches 256-512. After that point, the quality plateaus and gently decreases. We settle on $256$ dimensions as it exhibits a consistent behaviour across datasets and tasks. For overlapping community detection, $128$ and $256$ dimensions display comparable results; we opt for $128$ for the sake of efficiency. 

On the other hand, modularity is maximum at $16$ dimensions. This discrepancy between NMI and modularity reinforces once more the observation that the pure modularity optimization of models such as \dmon does not necessarily lead to superior quality.  
Finally, the results for other datasets follow a similar trend, confirming the robustness of \ourmethod. }



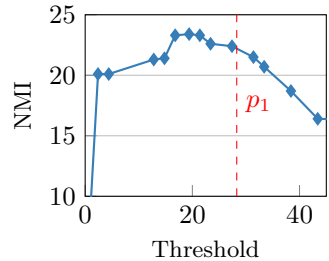
\begin{wrapfigure}{r}{.4\textwidth}
\vspace{-7mm}
\begin{tikzpicture}
\begin{axis}[
    ylabel={\small NMI},
    xlabel={\small Threshold},
    xtick pos=bottom,
    ytick pos=left,
    height=4.0cm,
    width=.4\textwidth,
    ymin=10,
    ymax=25,
    xmin=0,
    xmax=45,
    every axis plot/.append style={thick},
    xticklabel style={/pgf/number format/fixed},ymajorgrids]
]
\addplot[color=cycle2,mark=diamond*, mark size=2pt] table [x=Threshold, y=ONMI] {results/nmi-686-threshold.csv}; 
 \end{axis}
\draw [color=red, dashed] (2.0,0) -- node[right] {$p_1$}  (2.0,2.45);
\end{tikzpicture}
\caption{\dme{NMI vs. community assigment threshold; Fb-686 dataset.}}
\label{fig:nmi-epochs-threshold}
\vspace{-3mm}
\end{wrapfigure}

\dme{Figure~\ref{fig:nmi-epochs-threshold} reports the overlapping threshold {$p_1$} for the Fb-686 dataset as an example of a dataset with overlapping communities; \rev{we observe similar results in other datasets. 
The results indicate that there is a relatively broad range of values within $[0,40]$ in which our method performs well. A threshold $>22$ misses relevant community assignments, while a low value assigns every node to all communities. The choice $ 21$ corresponds to the earlier discussed \rev{setting $p_{1}$ (Figure~\ref{fig:nmi-epochs-threshold}, red line).}}}


\subsection{Ablation study}

Table~\ref{tbl:ablation}~in \revm{Section~\ref{sssec:theory}}\rev{ shows the results of the ablation study on three datasets for non-overlapping and four datasets for overlapping community detection. We experiment with a variant of our loss function in Equation~\ref{eq:objective} only with intra-cluster similarity (modularity), only with inter-cluster similarity (row-permuted modularity) and \ourmethod's loss.
In non-overlapping community detection, the intra-cluster similarity produces noisy communities. Yet, the results improve significantly with a combination of the two modularity scores, as the objective drives the model to discriminate true communities from noise. 
In overlapping community detection, the effect of the row-permuted modularity is more tangible and vindicates the choice of our contrastive loss showing a sensitive increase in performance when both similarities are introduced. Furthermore, the introduction of overlapping community probabilities in UCoDe effectively encourages the model to discover nodes belonging to multiple communities.
The results show that the combination of the intra-cluster and the inter-cluster similarity brings the largest benefit.}

\section{Conclusion}\label{sec:conclusion}
\ia{We propose \ourmethod, a new Graph Neural Network method for community detection in attributed graphs. \dmw{\ourmethod performs both overlapping and non-overlapping community detection, by virtue of a novel contrastive loss that maximizes a soft version of network modularity.
Our experimental assessment confirms that our method is expressive and overall superior in both overlapping and non-overlapping community detection tasks, exhibiting competitive performance in comparison with state-of-the-art methods designed for either one of the tasks.} 
}

\section*{Declarations}

\spara{Funding.} Atefeh Moradan is supported by the Innovationsfonden Denmark under the Grand Solutions project Hospital@Night. 

\spara{Conflicts of interest/Competing interests.} 
The authors have conflicts with 
\begin{itemize}
    \item Aarhus university (au.dk)
    \item Juelich research center (fz-juelich.de)
    \item Ira Assent: Co-author
    \item Ilaria Bordino: Recent collaborator
    \item Francesco Gullo: Recent collaborator
    \item Panagiotis Karras: Colleague
    \item Thomas Seidl: PhD advisor
\end{itemize}

\spara{Ethics approval.} Not applicable. 

\spara{Consent to participate.} The authors provide the appropriate consent to participate. 

\spara{Consent for publication} The authors provide the consent to publish the images in the manuscript. The data used in the publication is publicly available. We provide respective citations for each of the data sources. 

\spara{Availability of data and material.} The data and the code are available at  \href{https://github.com/AU-DIS/UCODE}{https://github.com/AU-DIS/UCODE}

\spara{Code availability}: The data and the code are available at \href{https://github.com/AU-DIS/UCODE}{https://github.com/AU-DIS/UCODE}

\spara{Authors' contributions}. 
\begin{itemize}
    \item Atefeh Moradan contributed to the concept, the experiments, the writing, and the algorithms. 
    \item Andrew Draganov contributed to the theory, part of the experiments, and writing. 
    \item Davide Mottin and Ira Assent contributed to the supervision, the writing, and the correction of the paper. 
\end{itemize}

\bibliography{bibliography}
\balance
\clearpage
\appendix
\revm{\section{Additional material}\label{sec:Appendix}
Here we introduce additional material for reproducing the experiments and support further the analyses in the paper.}

\revm{\subsection{Baselines}

\subsubsection{Non-overlapping}
\label{sssec:app:non-ovelapping}

We evaluate our method against the following established non-overlapping community detection methods: 
\begin{itemize}[leftmargin=*, noitemsep, topsep=0pt]
\item \textbf{\kmeans} clusters node attributes with the \kmeans++ algorithm~\cite{David2007kmeans}; we use the implementation in the scikit-learn package\footnote{\href{https://scikit-learn.org/stable/modules/generated/sklearn.cluster.KMeans.html}{https://scikit-learn.org/stable/modules/generated/sklearn.cluster.KMeans.html}}.
\item \rev{\textbf{\louvain}~\cite{blondel2008fast} is a heuristic method for modularity maximization; we use the implementation in the NetworkX library\footnote{\href{https://networkx.org/documentation/stable/reference/algorithms/generated/networkx.algorithms.community.louvain.louvain_communities.html}{https://networkx.org/}}} 
\item\textbf{\dgi}: Deep Graph Infomax (DGI)~\cite{velivckovic2018deep} is an unsupervised GNN model. After obtaining DGI node representations, \kmeans clusters these representations; we use the implementation from the authors\footnote{\href{https://github.com/PetarV-/DGI}{https://github.com/PetarV-/DGI}}
\item\textbf{\dmon}~\cite{tsitsulin2020graph} is a state-of-the-art community detection model that trains a shallow GCN to exclusively optimize graph modularity; we use the implementation from the authors \footnote{\href{https://github.com/google-research/google-research/tree/master/graph_embedding/dmon}{https://github.com/google-research/google-research/tree/master/graph\_embedding/dmon}}.
\item\textbf{\mincut}~\cite{bianchi2020spectral} is a graph pooling technique that trains a GNN with a min-cut loss similar to spectral clustering~\cite{shi2000normalized}; we use the pytorch implementation from \dmon~\cite{tsitsulin2020graph}. 
\dme{\item\textbf{\dcrn}~\cite{liu2022deep} is the most recent GNN  for non-overlapping community detection. \dcrn employs a combined objective and requires a pre-trained DFCN~\cite{tu2021deep} network to initialize the model embeddings; the communities are \kmeans clusters of the output embeddings. We downloaded the implementation from the authors\footnote{\href{https://github.com/yueliu1999/DCRN}{https://github.com/yueliu1999/DCRN}}, including the pre-trained networks. We tried to reproduce the experiments in the best of our capacity using the same version of the libraries, hyperparameters, and code, but the results were inconsistent with the ones reported in \cite{liu2022deep}. After a thorough investigation, we realized that the reported values must be the maximum NMI across epochs. However, in an unsupervised task, the ground-truth communities are unknown, hence the maximum NMI is unknown as well. 
Even considering the maximum value, the method does not attain the results declared in the paper. 
We, therefore, report results obtained using standard evaluation methodology, i.e., NMI for a fixed number of training epochs, consistent with the remainder of our experiments. 
}
\end{itemize}

\subsubsection{Overlapping}
\label{sssec:app:overlap}

We compare \ourmethod against the following baselines and state-of-the-art methods for overlapping community detection, \dme{including DMoN}:
\begin{itemize}[leftmargin=*, noitemsep, topsep=0pt]
\item\textbf{NOCD}~\cite{shchur2019overlapping} is a GCN based on the BigClam objective; we use the implementation from the authors\footnote{\href{https://github.com/shchur/overlapping-community-detection}{https://github.com/shchur/overlapping-community-detection}}
\rev{\item\textbf{COPRA}~\cite{gregory2010finding} discovers an arbitrary number of overlapping communities via label propagation; we use the implementation from the authors\footnote{\href{https://gregory.org/research/networks/copra/}{https://gregory.org/research/networks/copra/}}}
\rev{\item\textbf{CDE~\cite{CDE2018}} and \textbf{SNMF}~\cite{SNMF2010} \dme{employ non-negative matrix factorization to detect communities; results are from~\cite{shchur2019overlapping}}
\item \textbf{BigClam}~\cite{yang2013overlapping} finds overlapping communities optimizing \dme{the parameters of} a Bernoulli-Poisson model; results are from~\cite{shchur2019overlapping}.}
\end{itemize}
}

\revm{\subsection{Complete results for non-overlapping communities}

For completeness, Table~\ref{tbl:nmi-nonoverlapping} reports the NMI values and Table~\ref{tbl:f1-nonoverlapping} the F1 values for each method reported in Figure~\ref{fig:f1-nmi-nonoverlapping}.


\begin{table*}[!htb]
\vspace{-2mm}
\footnotesize
\setlength{\tabcolsep}{1pt}
\newcolumntype{C}{>{\centering\arraybackslash}X}
\newcolumntype{H}{>{\setbox0=\hbox\bgroup}c<{\egroup}@{}}
\newcolumntype{R}{>{\raggedleft\arraybackslash}X}
\begin{tabularx}{\textwidth}{p{1.4cm}r*{8}{R}}
\toprule
Dataset & \kmeans & \revc \louvain & \revc DCRN & \dgi & \mincut & \nocd & \dmon & \ourmethodk & \ourmethod \\ 
\midrule
\cora  & 15.0 & \revc 45.60 & \revc 33.3 & 55.4 & 37.1 & 45.8 & 46.3 & \uline{55.7}  &\textbf{57.4}  \\

\citse & 22.0 & \revc 32.7 & \revc 23.0 & \uline{42.5} & 23.1 &  23.4 & 31.4 &  \textbf{44.4}  & 41.0  \\ 

\pubmed & \textbf{31.0} & \revc 20.1 & \revc 0.0 & \uline{30.0} & 23.6 & 23.7  & 25.1 & 23.5 & 25.0 \\  

\amzph & 13.6 & \revc \uline{65.2} & - & 15.6 & - & 60.6 & 55.0 & 61.0  & \textbf{67.8} \\

\amzpc & 12.0 & \revc \textbf{52.3} & - & 11.8 & - & \uline{46.8} & 44.3 & 38.5 & 44.4 \\

\coacs & 33.8 & \revc 57.3 & \revc 63.3 & 67.5 & 68.1 & 73.6 & 67.7 &\textbf{78.4} & \uline{77.0}\\

\coaphy & 20.9 & \revc 45.7 & \revc 54.7 & 51.0 & 45.9 & 52.8  & 49.8 & \uline{55.3} &\textbf{57.5}\\

\bottomrule
\end{tabularx}
\caption{NMI for non-overlapping community detection. The best performer is highlighted in bold, and the second best is underlined.} 
\label{tbl:nmi-nonoverlapping}
\vspace{-4mm}
\end{table*}


\begin{table*}[htb]
\scriptsize
\setlength{\tabcolsep}{1pt}
\newcolumntype{C}{>{\centering\arraybackslash}X}
\newcolumntype{R}{>{\raggedleft\arraybackslash}X}
\newcolumntype{H}{>{\setbox0=\hbox\bgroup}c<{\egroup}@{}}
\begin{tabularx}{\textwidth}{p{1.4cm}r*{8}{C}}
\toprule
Dataset & \kmeans & \revc \louvain  & \revc \dcrn & \dgi & \mincut  & \nocd & \dmon & \ourmethodk &\ourmethod   \\ 
\midrule
\cora  & 44.1 & \revc 39.1 & \revc 47.8 & \textbf{63.4} & 40.0 & 40.6  & 49.6 & \uline{61.1} & 47.5 \\
\citse & 33.8 & \revc 20.0 & \revc 48.1 & \uline{55.5} & 30.7 & 27.1 & 44.2 & \textbf{55.9} & 52.7  \\ 
\pubmed  & 47.9 & \revc 21.1 & \revc 22.3 & \textbf{50.3} & 37.2 & 18.5 & 39.1 & \uline{49.3} & 36.1  \\  
\amzph & 23.4 & \revc \uline{63.2} & - & 41.2 & - & 56.7 & 57.9 & 52.9 & \textbf{66.2}  \\
\amzpc & 20.8 & \revc 38.8 & - & 42.7 & - & 38.2 & \uline{46.4} & 36.0 &\textbf{49.8} \\
\coacs & 41.2 & \revc 48.4 & \revc 42.3 & 59.3 & 58.8 & 60.2 & 58.0 &\textbf{79.7} &\uline{79.0} \\
\coaphy & 41.2 & \revc 38.6 & \revc 56.6 & 30.6 & 47.7 & 31.4  & 47.6 & \uline{48.5} & \textbf{49.0} \\
\bottomrule
\end{tabularx}
\caption{F1-result scores for non-overlapping community detection results on the seven real world data sets as summarized in Figure \ref{fig:f1-nmi-nonoverlapping}. The best performer is highlighted in bold, and the second best is underlined.
} 
\label{tbl:f1-nonoverlapping}
\vspace{6pt}
\end{table*}
}

\revm{\subsection{Statistical significance test}
\label{sssec:app:ttest}
We perform an individual two-sided t-test using NMI to compare each model with \ourmethod. The arrows in Table~\ref{tbl:ttest1-nonoverlapping} indicate a statistically significant difference (with p-value $<0.05$) compared to \ourmethod. The results demonstrate that in 85\% of the cases, \ourmethod is significantly better than the competitors.}


\begin{table*}[!htb]
\vspace{-2mm}
\footnotesize
\setlength{\tabcolsep}{1pt}
\newcolumntype{C}{>{\centering\arraybackslash}X}
\newcolumntype{H}{>{\setbox0=\hbox\bgroup}c<{\egroup}@{}}
\newcolumntype{R}{>{\raggedleft\arraybackslash}X}
\begin{tabularx}{\textwidth}{p{1.4cm}r*{8}{R}}
\toprule
Dataset & \cora & \revc \citse & \revc \pubmed & \amzph & \amzpc &\coacs & \coaphy \\ 
\midrule
\kmeans  & \win & \win & \lose& \win & \win & \win & \win  \\

\louvain  & \win & \win & \win& \win & \lose & \win & \win   \\ 
  
\dgi   & \win & \lose & \lose& \win & \win & \win & \win   \\ 

\mincut & \win & \win & \win& \win & \win & \win & \win   \\ 

\nocd & \win & \win & \win& \win & \lose & \win & \win   \\ 
\dmon & \win & \win & --& \win & -- & \win & \win   \\

\bottomrule
\end{tabularx}
\caption{Results of the t-tests using Normalized Mutual Information (NMI); p-value $<0.05$; the arrows indicate statistical significance; \win indicates that \ourmethod is significantly better than the competitor.} 
\label{tbl:ttest1-nonoverlapping}
\vspace{-4mm}
\end{table*}


  




\revm{\subsection{Extended sensitivity analysis}
 Figure~\ref{fig:embedding-epochs-threshold-amazph} extends the analysis in Section~\ref{ssec:tuning}
 to the \citse and \amzph datasets. The results consistently indicate 256 as an optimal embedding dimension for the intermediate layer. 


\begin{figure}[!h]
\begin{tikzpicture}
\centering
\begin{groupplot}[group style={
                      group name=myplot,
                      group size= 2 by 2, horizontal sep=1.5cm,
                      vertical sep=1cm},
                      height=4.0cm,
                      width=.49\textwidth,
                      title style={align=center, at={(1.2,1.2)},anchor=north},
                      ymin=0,
                      ymax=0.8,
                      every axis plot/.append style={thick},
                      xticklabel style={/pgf/number format/fixed},ymajorgrids]
\nextgroupplot[
	ylabel={\small NMI},
	ymin=35,
	ymax=50,
	xmin=8, 
	xmax=512,
	xmode=log,
	xtick={8,16,32,64,128,256,512},
	log ticks with fixed point,
 	title = {\textbf{\citse}},
]
\addplot[color=cycle2,mark=diamond*, mark size=2pt] table [x=Dimension, y=NMI, y expr=\thisrowno{1}*100] {results/nmi-dim-citeseer.csv};
\nextgroupplot[
	ylabel={\small Modularity ($Q$)},
	ymin=65,
	ymax=85,
	xmin=8, 
	xmax=512,
	xmode=log,
	y label style={at={(-0.12,0.5)}},
	xtick={8,16,32,64,128,256,512},
	log ticks with fixed point,
]
\addplot[color=cycle2,mark=diamond*, mark size=2pt] table [x=Dimension, y=Modularity] {results/mod-dim-citeseer.csv};

\nextgroupplot[
	xlabel={\small Embedding dimension},
	ylabel={\small NMI},
	ymin=45,
	ymax=70,
	xmin=8, 
	xmax=512,
	xmode=log,
	xtick={8,16,32,64,128,256,512},
	log ticks with fixed point,
 	title = {\textbf{\amzph}},
]
\addplot[color=cycle2,mark=diamond*, mark size=2pt] table [x=Dimension, y=NMI, y expr=\thisrowno{1}*100] {results/nmi-dim-amazon_photo.csv};
\nextgroupplot[
	xlabel={\small Embedding dimension},
	ylabel={\small Modularity ($Q$)},
	ymin=55,
	ymax=75,
	xmin=8, 
	xmax=512,
	xmode=log,
	y label style={at={(-0.12,0.5)}},
	xtick={8,16,32,64,128,256,512},
	log ticks with fixed point,
]
\addplot[color=cycle2,mark=diamond*, mark size=2pt] table [x=Dimension, y=Modularity] {results/mod-dim-amazon_photo.csv};
\end{groupplot}
\end{tikzpicture}
\caption{Impact of the embedding dimension on NMI and modularity for non-overlapping community detection; \amzph and \citse datasets.}
\label{fig:embedding-epochs-threshold-amazph}
\end{figure}
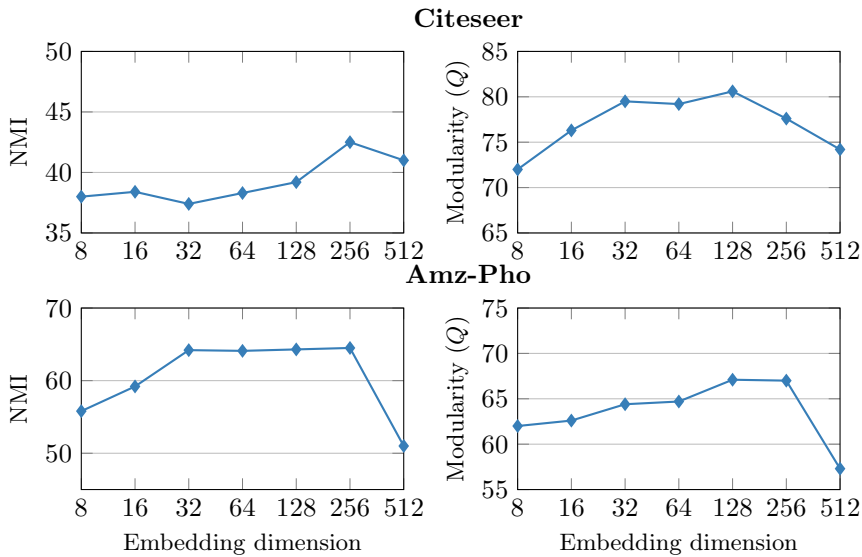

We additionally present the analysis of the loss function for \citse and \amzph in Figure~\ref{fig:tuning_extra}. The loss exhibits a steady increasing behaviour, stabilizing around 100 epochs as experienced in Section~\ref{ssec:tuning}. Interestingly, \dmon's performance is more fluctuating in \amzph as opposed to other datasets.    
}
\revm{\begin{figure*}[!h]
\small
\vspace{-4mm}
\begin{tikzpicture}
\begin{groupplot}[group style={
                      group name=myplot,
                      group size= 3 by 2, horizontal sep=1.1cm,
                      vertical sep=1cm},
                      title style={align=center, at={(1.8,1.2)},anchor=north},    
                      height=3.5cm,
                      width=0.36\linewidth ,
                      ymin=0,
                      ymax=0.8,
                      every axis plot/.append style={thick},
                      xticklabel style={/pgf/number format/fixed},ymajorgrids]
                      
\nextgroupplot[
	ylabel={\small Loss},
	ymin=0.6,
	ymax=0.7,
	xmin=0, 
	xmax=500,
        title={\textbf{\citse}},
]
\addplot[color=cycle2] table [x=Epoch, y=Loss ,col sep=comma]{results/loss_citeseer.csv};

\nextgroupplot[
	ylabel={\small NMI},
	y label style={at={(-0.14,0.5)}},
	legend entries={ DMoN,\ourmethod },
	legend pos=south east,
    xtick pos=bottom,
    ytick pos=left,
	ymin=0,
	ymax=50,
	xmin=0, 
	xmax=500,
]
\addplot[color=cycle1] table [x=epoch, y=nmi,  y expr=\thisrowno{1}*100, col sep=comma] {results/nmi_modul_citeseer_dmon.csv};
\addplot[color=cycle2] table [x=Epoch, y=NMI , y expr=\thisrowno{2}*100, col sep=comma] {results/loss_nmi_modularity_citeseer_ucode.csv};

\nextgroupplot[
	ylabel={\small Modularity ($Q$)},
	y label style={at={(-0.14,0.5)}},
	legend entries={ DMoN, \ourmethod },
	legend pos=south east,
    xtick pos=bottom,
    ytick pos=left,
	ymin=0,
	ymax=85,
	xmin=0, 
	xmax=500,
]
\addplot[color=cycle1] table [x=epoch, y=modularity,  y expr=\thisrowno{2}*100, col sep=comma] {results/nmi_modul_citeseer_dmon.csv};
\addplot[color=cycle2] table [x=Epoch, y=Modularity , y expr=\thisrowno{3}*100, col sep=comma] {results/loss_nmi_modularity_citeseer_ucode.csv};
                      
\nextgroupplot[
	ylabel={\small Loss},
	xlabel={Epochs},
	ymin=0.6,
	ymax=0.7,
	xmin=0, 
	xmax=500,
        title={\textbf{\amzph}},
]
\addplot[color=cycle2] table [x=Epoch, y=Loss ,col sep=comma]{results/loss_amazon_photo.csv};

\nextgroupplot[
	ylabel={\small NMI},
	xlabel={Epochs},
	y label style={at={(-0.14,0.5)}},
	legend entries={ DMoN,\ourmethod },
	legend pos=south east,
    xtick pos=bottom,
    ytick pos=left,
	ymin=0,
	ymax=75,
	xmin=0, 
	xmax=500,
]
\addplot[color=cycle1] table [x=epoch, y=nmi,  y expr=\thisrowno{1}*100, col sep=comma] {results/nmi_modul_amazonphoto_dmon.csv};
\addplot[color=cycle2] table [x=Epoch, y=NMI , y expr=\thisrowno{2}*100, col sep=comma] {results/loss_nmi_modularity_amazon_photo_ucode.csv};

\nextgroupplot[
	ylabel={\small Modularity ($Q$)},
	xlabel={Epochs},
	y label style={at={(-0.14,0.5)}},
	legend entries={ DMoN, \ourmethod },
	legend pos=south east,
    xtick pos=bottom,
    ytick pos=left,
	ymin=0,
	ymax=80,
	xmin=0, 
	xmax=500,
]
\addplot[color=cycle1] table [x=epoch, y=modularity,  y expr=\thisrowno{2}*100, col sep=comma] {results/nmi_modul_amazonphoto_dmon.csv};
\addplot[color=cycle2] table [x=Epoch, y=Modularity , y expr=\thisrowno{3}*100, col sep=comma] {results/loss_nmi_modularity_amazon_photo_ucode.csv};
\end{groupplot}

\end{tikzpicture}
\caption{\revm{Training \ourmethod that quickly minimizes the loss (left); NMI increases steadily and achieves $9\%$ ($19\%$ in \amzph) higher value than DMoN (center); \ourmethod gradually outperforms DMoN's modularity (right).}}
\vspace{-3mm}
\label{fig:tuning_extra}
\end{figure*}
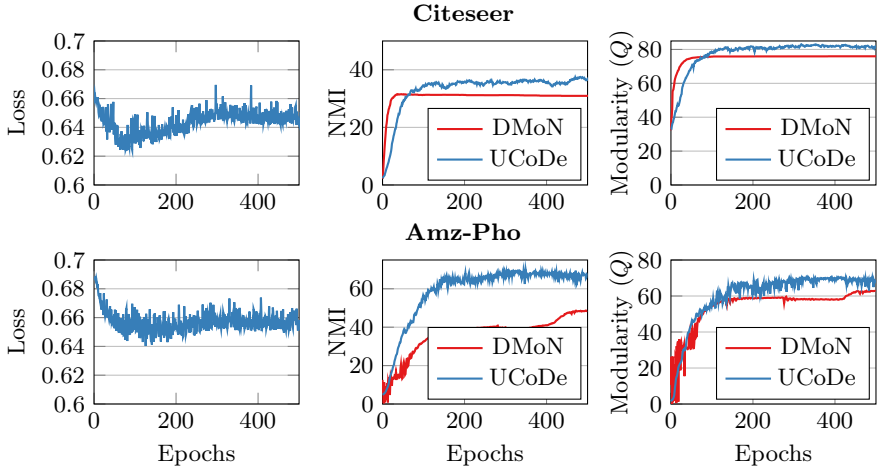}

\end{document}